\documentclass[11pt]{article}
\usepackage[margin=1.0in]{geometry} 
\usepackage{color}
\usepackage{fullpage}
\usepackage{hyperref} 
\definecolor{darkred}{rgb}{0.8,0,0}
\definecolor{darkgreen}{rgb}{.0,.8,.0}
\hypersetup{ colorlinks=true, linkcolor=blue, citecolor=red } 
\usepackage{amsmath, amsthm, amssymb,amsfonts}
\usepackage{epsfig}
\usepackage{graphicx}
\newcommand{\tkim}{\tilde k_{1\mu}}
\newcommand{\tkin}{\tilde k_{1\nu}}
\newcommand{\tkir}{\tilde k_{1\rho}}
\newcommand{\tkis}{\tilde k_{1\sigma}}

\newcommand{\hf}{\frac{1}{2}}

\newcommand{\xn}{x_{n}}

\newcommand{\kim}{ k_{1\mu}}                                      
\newcommand{\kom}{ k_{0\mu}}                                      
\newcommand{\ki}{ k_{1}}
\newcommand{\yi}{ Y_{1}}
\newcommand{\yib}{ Y_{\bar 1}}

\newcommand{\kn}{ k_{n}}

\newcommand{\kt}{ k_{2}}                                             
\newcommand{\ko}{ k_{0}}                                             
\newcommand{\yim}{ Y_{1}^{\mu}}                                      
\newcommand{\yin}{ Y_{1}^{\nu}}                                      
\newcommand{\kin}{ k_{1\nu}}  
\newcommand{\kir}{ k_{1\rho}} 
\newcommand{\kis}{ k_{1\sigma}}                                    
\newcommand{\kon}{ k_{0\nu}}
\newcommand{\kor}{ k_{0\rho}}
\newcommand{\kos}{ k_{0\sigma}}

\newcommand{\p}{\partial}                                           
\newcommand{\pp}{\partial ^{2}}

\newcommand{\li}{ \lambda_{1}} 
\newcommand{\lib}{ \lambda_{\bar 1}}

\newcommand{\al}{\alpha }

\newcommand{\lan}{\langle}
\newcommand{\ran}{\rangle}

\newcommand{\kimb}{\mbox {$ {k_{\bar1\mu}}$}}                               
  
\newcommand{\kirb}{\mbox {$ {k_{\bar1\rho}}$}}                               
\newcommand{\kisb}{\mbox {$ {k_{\bar1\sigma}}$}}

\newcommand{\kib}{\mbox {$ {k_{\bar1}}$}}

\newcommand{\qo}{ q_{0}}

\newcommand{\la}{ \lambda }                                           
\newcommand{\be}{\begin{equation}}                                             
\newcommand{\br}{\begin{eqnarray}}                                             
\newcommand{\ee}{\end{equation}}                                               
\newcommand{\er}{\end{eqnarray}}

\newcommand{\ppp}{\mbox {$ \partial ^{3}$}}

\begin{document}
\title{
\hfill\parbox{4cm}{\normalsize IMSC/2014/12/13\\
}\\
\vspace{2cm}
Background Independence, Gauge Invariance and Equations of Motion for Closed String Modes
%\thanks{}%
}
\author{B. Sathiapalan\\ {\em                                                  
Institute of Mathematical Sciences}\\{\em Taramani                     
}\\{\em Chennai, India 600113}}                                     
\maketitle                                                                     
\begin{abstract}   
In an earlier paper [arXiv:1408.0484] gauge invariant and background covariant equations for closed string modes were obtained from the exact renormalization group equation of the world sheet theory. The background metric (but not the physical metric) had to be flat and hence the method was not manifestly background independent. In this paper the restrictions on the background metric are relaxed. A simple prescription for the map from loop variables to space time fields is given whereby for arbitrary backgrounds the equations are generally covariant and gauge invariant.  Extra terms involving couplings of the curvature tensor to (derivatives of) the Stueckelberg fields have to be added.  The background metric is chosen to be the physical metric without any restrictions.  This method thus gives manifestly background independent gauge invariant and general covariant equations of motion for both open and closed string modes.  
 \end{abstract}  
 
\newpage 
\tableofcontents 
                                                             
\newpage                                                                       
\section{Introduction} 

This paper describes a manifestly background independent generalization of the results of \cite{BSGIC,BSERGclosed}. 
In \cite{BSGIC,BSERGclosed} a method of writing down gauge invariant and background covariant equations for the closed string was described. This is a generalization of earlier work on the renormalization group (RG) approach to obtaining string equations of motion [\cite{L}-\cite{T}]. This made use of loop variables, and was first applied to free open strings in \cite{BSLV}. Gauge invariant equations obtained were similar to those obtained using BRST string field theory [\cite{SZ} -\cite{BZ}]. The interacting open string was described in \cite{BSERGopen1, BSERGopen2}. This involved using the exact renormalization group (ERG) \cite{WK,W,W2,P} \footnote{See [\cite{BB1}-\cite{IIS}] for more recent interesting results on ERG.} of the world sheet theory, written in terms of loop variables. The result is a gauge invariant interacting field theory for the modes of the open string.

For closed strings, in \cite{BSGIC,BSERGclosed}, a world sheet theory was written using a {\em background} space-time (i.e. target space) metric. This was over and above   a dynamical  {\em physical} space time metric. The kinetic term and interaction term separately were made invariant under background coordinate transformations i.e. coordinate transformations that also transformed the background metric. An exact renormalization group (ERG)  for the world sheet theory was written down. 
As a result the equations obtained are manifestly covariant under background transformations. The full world sheet action was written down so that it did not have a dependence on the background metric - the background metric dependence canceled between the kinetic and interaction terms.  Thus since the action does not actually depend on the background metric it is expected that the {\em solutions} of the full set of equations will be independent of the background metric. In such a situation background covariance becomes the usual general covariance. 

These equations turn out to be also invariant under the  gauge transformations involving the massive string modes {\em only} when the curvature associated with the background metric is zero. Thus while the physical metric is arbitrary, the background metric has to be flat. In this sense the method is not {\em manifestly} background independent and is tied to flat backgrounds. This also precludes the convenient option of choosing the background metric to be the physical metric - which is useful in background field formalisms [\cite{Abbott}-\cite{GNW}]. 

In this paper we show how to get around this restriction of zero curvature. The main new result of this paper is a new prescription for mapping loop variable expressions to expressions involving space-time fields, that guarantees consistency with gauge invariance. It is applicable to both closed and open strings. The equations of motion obtained in this paper are both gauge invariant and generally covariant in an arbitrary background. The method is thus {\em manifestly} background independent. \footnote{ Background independence in the BRST approach is described in [\cite{Wi}- \cite{KMM}].}

The earlier restriction to flat backgrounds can be understood in a very simple way: Consider a loop variable expression $\kom  \kir \kis$.\footnote{Loop variables are described in \cite{BSGIC,BSERGclosed,BSERGopen1,BSERGopen2}. This paper should be read in conjunction with these papers.}
In flat space this maps to the space time field expression as
\be
\lan \kom  \kir \kis \ran = \p_\mu S_{11\rho \sigma}
\ee
In curved space time we work in the Riemann Normal Coordinate (RNC) system and interpret $i \kom \approx {\p\over \p \bar Y^\mu}$ where $\bar Y^\mu$ are RNC's. We extract $\bar Y^\mu$ dependence of  a field using Taylor expansions \cite{Pet}. Thus if $x_0$ is the origin of the RNC and $x$ a general point, then if we let $y^\mu$ be defined by $x(y)^\mu =x_0^\mu +y^\mu$ in the RNC, then the following is the Taylor series expansion for a tensor {\em in the RNC} written in terms of objects that are manifestly tensors at the origin. (To change to other coordinate systems one can transform the LHS as a tensor at $x$ and the RHS as a sum of tensors at $x_0$. Thus they transform differently
and one must compensate for this.)

\[
W_{\al _1 ....\al _p}(x) 
= W_{\al _1 ....\al _p}(x_0) ~+~
W_{\al _1 ....\al _p , \mu}(x_0)y^\mu ~+~
\]
\[
{1\over 2!}\{W_{\al _1 ....\al _p ,\mu \nu}(x_0) 
~-~{1\over 3}
\sum _{k=1}^p R^\beta _{~\mu \al _k \nu}(x_0) 
W_{\al _1 ..\al _{k-1}\beta \al _{k+1}..\al _p}(x_0)\}
 y^\mu y^\nu
~+~
\]
\[   
{1\over 3!}\{W_{\al _1 ....\al _p ,\mu \nu \rho}(x_0) - 
\sum _{k=1}^p R^\beta _{~\mu \al _k \nu}(x_0) 
W_{\al _1 ..\al _{k-1}\beta \al _{k+1}..\al _p, \rho}(x_0)
\]
\be \label{Taylor}
-
{1\over 2}\sum _{k=1}^p R^\beta _{~\mu \al _k \nu ,\rho }(x_0) 
W_{\al _1 ..\al _{k-1}\beta \al _{k+1}..\al _p}(x_0)\}y^\mu y^\nu y^\rho +...
\ee
The commas denote covariant derivatives.

Thus $\p_\mu S_{11\rho\sigma}$ in RNC can be written as $\nabla_\mu S_{11\rho\sigma}$. Thus
\be  \label{map1}
\lan \kom  \kir \kis \ran = \nabla_\mu S_{11\rho \sigma}
\ee

Now consider the gauge transformation of $ \kir \kis \rightarrow  \kir \kis +  k _{0\rho} \la_1 \kis +  k _{0\sigma} \la _1 \kir$. This becomes
\be   \label{gauge1}
\delta S_{11\rho\sigma}=\lan k _{0\rho} \la_1 \kis +  k _{0\sigma} \la _1 \kir\ran = \nabla _\rho \Lambda_{11\sigma}+\nabla_\sigma \Lambda_{11\rho}
\ee
Now consider the map of the gauge transformation of $\kom  \kir \kis$, $\kom(k _{0\rho} \la_1 \kis +  k _{0\sigma} \la _1 \kir)$. It becomes,
using the Taylor expansion \eqref{Taylor}:
\be    \label{map2}
\lan \kom(k _{0\rho} \la_1 \kis +  k _{0\sigma} \la _1 \kir)\ran = \nabla _\mu(\nabla _\rho \Lambda_{11\sigma}+\nabla_\sigma \Lambda_{11\rho}) - {2\over 3} (R^\beta_{~\rho \mu \sigma} + R^\beta _{~ \sigma \mu \rho})\Lambda_{11\beta}
\ee
The extra curvature coupling is required because of the non commutativity of covariant derivatives. We see that \eqref{map2} is not obtained from \eqref{map1} by a gauge transformation \eqref{gauge1}.  Thus expressions that are gauge invariant when written in terms of loop variables are no longer so under the naive map to space time fields - we need to take into account the non commutativity of covariant derivatives. If curvature is zero, this problem does not arise. 

 A solution to this problem is described in this paper. A modified prescription is given for the map to space-time fields that does not suffer from the problem described above. This allows arbitrary curved background metrics.

The covariantization of the world sheet action is also done in a different manner from \cite{BSGIC}. We show that by a symmetrization prescription, i.e. symmetrize all covariant derivatives, the world sheet action in Riemann normal coordinates, can be written easily in a covariant form.

For completeness let us explain a  difference with \cite{BSGIC,BSERGclosed}. There a more elaborate construction was required. Background covariant derivatives were introduced in the usual way. But since the action is not allowed to depend on this arbitrary background, the dependence introduced by covariantizing derivatives, was canceled by adding further terms in the action. These terms were
then absorbed into field redefinitions of the space-time (target space) fields describing the world sheet action. All this can be avoided if the background metric can be set equal to the metric. This was not possible in  \cite{BSGIC,BSERGclosed} because the background metric was constrained to be flat. In this paper this restriction is removed, so we can just covariantize the world sheet action, and let it depend on the physical metric.

We will explain the procedure and work out a few illustrative examples for both open and closed string equations. Note that in \cite{BSERGopen1,BSERGopen2,BSERGclosed, BSGIC}gauge invariant loop variable expressions are available along with their maps to space time in flat space. We will simply use
them and describe a consistent map to space time fields in curved space-time in Section 2. In section 3 we work out some examples.  Section 4 describes the covariantization of the world sheet action. Section 5 contains conclusions.

\section{Prescription}

\subsection{Tilde Variables}

In this section we give a simple prescription that guarantees that the map from loop variables to space time is consistent with gauge transformations. The crucial observation is the following. The naive map needs to be modified to include curvature term {\em whenever
extra derivatives appear}. Thus for example $\kom \kir \kis$ is mapped to $\nabla _\mu S_{11\rho \sigma}$. But a gauge transformation
produces an extra derivative as in $\kir\to k_{0\rho}\la_1$ and we have seen that $\kom \kor$ is not simply mapped to $\nabla _\mu \nabla_\rho$ because covariant derivatives do not commute.  Thus whenever a gauge transformation produces a derivative term there is a problem.

We solve this problem by rewriting the loop variable expression in such a way that no extra derivative terms appear in any gauge transformation - all derivatives lurking in gauge transformations will be made manifest right away. This can be implemented as follows:
Define
\be
 k_{n\mu} =\tilde k_{n\mu} + y_n\kom
\ee
where $y_n \to y_n + \la _n$ under a gauge transformation. $y_n$ have been defined earlier \cite{BSERGopen2}:
\be
\sum_{n=0} q_n t^{-n} = q_0 e^{\sum_{m=1} y_m t^{-m}}
\ee
Gauge transformation of $k_{n\mu}$ is given by
\be
k_{n\mu} \to k_{n\mu} + \la_1 k_{n-1\mu} + \la_2 k_{n-2\mu} +.....\la_{n-1}\kim + \la _n \kom
\ee 
Thus $\tilde k_{n\mu}$ satisfies a gauge transformation rule
\be
\tilde k_{n\mu} \to \tilde k_{n\mu} +   \la_1 k_{n-1\mu} + \la_2 k_{n-2\mu} +.....\la_{n-1}\kim
\ee
The crucial point is that no derivatives are involved. Thus we will rewrite our loop variable expression in terms of tilde variables.
Thus all the $\ko$ lurking in the gauge transformation are made manifest.  Then  the expression involving tilde variables are mapped to space time fields. This map is guaranteed to be consistent with gauge transformations. A field redefinition can now be made to get back to the original space-time fields. We illustrate this procedure below.

\subsection{Illustration of Procedure}

Consider the level 2 field
\be
\lan  \kim \kin \ran = S_{11\mu \nu}
\ee 
Let us define 
\be
 \kim = \tilde k_{1\mu} + y_1 \kom
\ee
Then
\be
\kim \kin =   \tilde k_{1\mu}  \tilde k_{1\nu} +  \kom y_1 \tilde k_{1\nu} + \kon y_1  \tilde k_{1\mu} + y_1^2 \kom \kon
\ee
Define 
%\footnote{The powers of $\qo$ are put in to conform with earlier notation where $q_n$ were used in defining fields.}

\br   \label{tilde}
\lan  \tilde k_{1\mu}  \tilde k_{1\nu} \ran &=& \tilde S_{11\mu\nu}\nonumber \\
\lan y_1 \tilde k_{1\nu}\ran &=& {\tilde S_{11\nu}}\nonumber \\
\lan y_1^2\ran &=& {\tilde S_{11}}
\er
 We then obtain
 \be   \label{FD1}
 S_{11\mu\nu} = \tilde S_{11\mu\nu} + \nabla _{(\mu} {\tilde S_{11\nu)}} + \nabla_\mu\nabla_\nu {\tilde S_{11}}
\ee
and
\be   \label{FD2}
 \tilde S_{11\mu} ={S_{11\mu}\over \qo} -{1\over \qo^2} \nabla_\mu S_{11}~~~~\tilde S_{11}= {S_{11}\over \qo^2}
\ee 
where 
\be
\lan q_1 \kim \ran = S_{11\mu} ;~~~~~\lan q_1^2\ran = S_{11}
\ee
and also $q_1=y_1 q_0$.

Let us turn to the gauge transformation laws for these fields: Using $\kim \to \kim + \li \kom$, $y_1 \to y_1+\li$, we obtain
\br  \label{GT1}
\delta \tilde S_{11\mu\nu} &=& 0 \nonumber \\
\delta \tilde S_{11\mu}&=& \lan \li \tilde k_{1\mu} \ran =\tilde \Lambda_{11\mu}\nonumber \\
\delta \tilde S_{11} &=& 2\lan y_1\li \ran = 2 \tilde \Lambda_{11}
\er
If we define 
\be
\lan \li \kim \ran = \Lambda_{11\mu} = \lan \li \tilde k_{1\mu} + \li y_1 \kom \ran =\tilde \Lambda _{11\mu} + \nabla _\mu \tilde \Lambda_{11}
\ee
we see that this is identical to
\br   \label{GT2}
\delta S_{11\mu\nu} &=& \nabla_{(\mu}\Lambda_{11\nu)}\nonumber \\
\delta S_{11\mu}&=& \nabla _\mu \Lambda_{11} + \qo \Lambda_{11\mu}\nonumber \\
\delta S_{11} &=& 2 \Lambda_{11}\qo
\er 
as described in earlier papers \cite{BSERGopen1,BSERGopen2,BSERGclosed}.

Now consider a loop variable expression 
\be  \label{LV1}
\kom \kir \kis
\ee
In terms of tilde variables it is
\be  \label{LV2}
\kom (\tilde k_{1\rho} \tilde k_{1\sigma} + \kor y_1  \tilde k_{1\sigma}+\kos  y_1 \tilde k_{1\rho} + \kor \kos y_1^2)
\ee
Mapping to space-time fields, keeping in mind \eqref{Taylor}, we get
\be  \label{ST1}
\nabla_\mu \tilde S_{11\rho\sigma} + \nabla _\mu \nabla_\rho \tilde S_{11\sigma}+ \nabla _\mu \nabla_\sigma \tilde S_{11\rho} +{2\over 3}(R^\beta _{~\rho \mu \sigma} +R^\beta _{~\sigma \mu \rho}) \tilde S_{11\beta} + \nabla _\mu \nabla _\rho \nabla_\sigma \tilde S_{11} + {1\over 3}(R^\beta_{~\rho \mu \sigma} +R^\beta_{~\sigma \mu \rho})\nabla_\beta \tilde S_{11}
\ee

Since the gauge transformation of the loop variable expression \eqref{LV2} does not produce any extra $\kom$ its space-time map is guaranteed to coincide with the gauge transformation of \eqref{ST1}. Thus we have an internally self consistent prescription for mapping to space-time fields.  

Let us now re-express \eqref{ST1} in terms of the original space-time fields without tildes:
using the field redefinitions \eqref{FD1},\eqref{FD2}. The three and two derivative term cancels (as expected) and we find
\be \label{ST2}
\nabla_\mu S_{11\rho\sigma} + {2\over 3} (R^\beta_{~\rho \mu \sigma} +R^\beta_{~\sigma \mu \rho})[{S_{11\beta}\over \qo} - \hf \nabla _\beta {S_{11}\over \qo^2}]
\ee

Now we can compare the gauge transforms of \eqref{LV1} and \eqref{ST2} to verify that they agree. The gauge variation of \eqref{LV1} is
\be  \label{GT3}
\kom (\kor \li \kis + \kos \li \kir)
\ee
Mapping \eqref{GT3} to space-time fields gives
\be
\nabla_\mu (\nabla_\rho \Lambda_{11\sigma} +\nabla_\sigma \Lambda_{11\rho}) + {2\over 3}(R^\beta _{~\rho \mu \sigma} +R^\beta _{~\sigma \mu \rho})\Lambda_{11\beta}
\ee
The gauge variation of \eqref{ST2} is (using \eqref{GT2}) seen to be the same as above.

Let us summarize the logic of the prescription: The map from loop variables to space-time fields becomes complicated in curved space-time due to the non commutativity of covariant derivatives. If a gauge transformation produces an extra derivative then this map produces extra curvature terms. By making explicit all the potential derivatives in the beginning, we ensure that no derivatives are generated during the gauge transformation. Thus no additional curvature couplings are generated. Thus the gauge transformation of an expression in loop variables, gives the same result whether we do the gauge transformation first and then map to space time or the other way around. 

\section{Example}
We turn to the actual equations in string theory at level 2  for open strings and level 4 in closed strings \footnote{The graviton case was worked out in detail in \cite{BSGIC} and is not modified by the new prescription introduced in the present paper, which is relevant only for the massive modes.}. The prescription for mapping from loop variables to space-time fields expressions is the same for both open and closed strings. We will work out some of the terms in the equations using the above prescription. We will not however work out all the terms since that is a tedious exercise and not very illuminating. The detailed equations are presumably only required when one attempts to solve the system - which is not the subject of this paper.

\subsection{Free Open strings: Level 2}

The free equation of motion is \cite{BSLV,BSERGopen1,BSERGopen2}:
\[
-\ki.\ki i\ko.D_2Y -\hf \ki.\ki (i\ko.D_1Y)^2 -\ki.\ko (\ki.D_1Y)(\ko.D_1Y) 
\]
\be
+\hf \ko^2 (\ki.D_1Y)^2 -\ko^2i\kt.D_2Y + i\ki.\ko(\ki.D_2Y) +\kt.\ko i\ko.D_2Y =0
\ee
The coefficient of $\yim \yin$ is
\be	\label{EOMspin2}
\ko^2 \kim \kin - \ko .\ki k_{1(\mu }k_{\nu)0} + \ki.\ki \kom \kon =0
\ee 

This is easily seen to be gauge invariant under $\kim \to \kim + \li \kom$. 
Dimensional reduction gives a massive field with EOM:
\be  \label{EOMspin2mass}
(\ko^2+\qo^2) \kim \kin -( \ko .\ki+ \qo q_1) k_{1(\mu }k_{\nu)0} + (\ki.\ki+q_1q_1) \kom \kon =0
\ee

This has to be mapped to space time fields.
The method was explained in the last section. There are three steps involved:
\begin{enumerate}
\item
Change to tilde loop variables
\item
Map to tilde space time fields. As explained earlier, with this map, gauge transformation is completely well defined in curved space because the gauge transformation of the tilde fields do not bring in derivatives.
\item
Rewrite the tilde fields in terms of ordinary fields. This is a simple field redefinition and does not modify the gauge transformation.
\end{enumerate}
At the end of this three step procedure we have a map from loop variables to space time fields that gives the correct gauge transformation (where correct is defined by what is obtained by mapping the gauge transformed loop variable expression to space time). This ensures that expressions that are gauge invariant in terms of loop variables continue to be gauge invariant in temrs of space time fields.

Let us apply this procedure to a general loop variable expression 
\be  \label{genterm}
 \kom \kon \kir\kis
\ee
 All the terms in \eqref{EOMspin2} can be obtained from this by contractions:

{\bf Step 1}

We let $\kim = \tilde k_{1\mu} + y_1 \kom$. Then \eqref{genterm} becomes
\be \label{step1}
\kom \kon \kir\kis=\kom \kon (\tkir \tkis + \tkir y_1 \kos + \kor y_1 \tkis +y_1^2 \kor \kos)
\ee
{\bf Step2}

Let us consider each term in turn and map to space time fields, using the definitions \eqref{tilde} and the expressions for Taylor expansion given in the Appendix.  We get
\br \label{step2}
\lan \kom \kon \tkir \tkis \ran &=& \hf[ \nabla _{(\mu }\nabla _{\nu)}\tilde S_{11\rho \sigma} + {1\over 3} R^\beta_{~ (\mu \nu) \rho}\tilde S_{11\beta \sigma} + {1\over 3}R^\beta_{~ (\mu \nu) \sigma}\tilde S_{11 \rho\beta}]\nonumber \\
\lan \kom \kon \kos \tkir y_1 \ran &=& {1\over 6}[\nabla_{(\mu}\nabla_\nu \nabla_{\sigma)} \tilde S_{11\rho}-R^\beta_{~(\mu |\rho|\nu } \nabla_{\rho)} \tilde S_{11\beta}- \hf \nabla_{(\sigma} R^\beta_{~\mu |\rho |\nu)}\tilde S_{11\beta}]\nonumber \\
\lan \kom \kon \kor \kos y_1^2\ran &=& {1\over 4!} \nabla_{(\mu} \nabla_\nu \nabla_\rho \nabla_{\sigma)} \tilde S_{11}
\er

The gauge transformations of the tilde variables are given in \eqref{GT1} and do not involve derivatives. Thus in \eqref{step2} the gauge transformations of the LHS and RHS are guaranteed to agree. Now we proceed to Step 3:

{\bf Step 3}

Now we can redefine fields in terms of the original fields using \eqref{FD1} and \eqref{FD2}:
\br   \label{step3}
\lan \kom \kon \tkir \tkis \ran &=& \hf \Big( \nabla _{(\mu }\nabla _{\nu)}[ S_{11\rho \sigma}- {\nabla_{(\rho} S_{11\sigma)}\over \qo^2}+ {\nabla_\rho \nabla_\sigma S_{11}\over \qo^2}] + \nonumber \\& &{1\over 3} R^\beta_{~ (\mu \nu) \rho}[ S_{11\beta \sigma}- {\nabla_{(\beta} S_{11\sigma)}\over \qo^2}+ {\nabla_\beta \nabla_\sigma S_{11}\over \qo^2}] + {1\over 3}R^\beta_{~ (\mu \nu) \sigma}[ S_{11\rho \beta}- {\nabla_{(\rho} S_{11\beta)}\over \qo^2}+ {\nabla_\rho \nabla_\beta S_{11}\over \qo^2}]\Big)\nonumber \\
\lan \kom \kon \kos \tkir y_1 \ran &=& {1\over 6}\Big(\nabla_{(\mu}\nabla_\nu \nabla_{\rho)} [ {S_{11\sigma}\over \qo} - {\nabla_\sigma S_{11}\over \qo^2}]-R^\beta_{~(\mu |\sigma|\nu } \nabla_{\rho)} [ {S_{11\beta}\over \qo} - {\nabla_\beta S_{11} \over \qo^2}]- \hf \nabla_{(\rho} R^\beta_{~\mu |\sigma |\nu)}[ {S_{11\beta}\over \qo} - {\nabla_\beta S_{11}\over \qo^2}]\Big)\nonumber \\
\lan \kom \kon \kor \kos y_1^2\ran &=& {1\over 4!} \nabla_{(\mu} \nabla_\nu \nabla_\rho \nabla_{\sigma)} { S_{11}\over \qo^2}
\er
Once again the gauge transformations of the combination of original fields, being identical to that of the tilde fields, the map from loop variables to space time fields continues to be well defined for gauge transformations also.
If we substitute \eqref{step3} in \eqref{step1} we find that all the higher derivative terms independent of curvature cancel as they should and give the expected flat space term: $\nabla_\mu \nabla_\nu S_{11\rho\sigma}$. Also in curved space-time it has the structure
\be    \label{op-step3}
\underbrace{\nabla_\mu \nabla_\nu S_{11\rho\sigma} +{1\over 3}(R^\beta _{~\nu \mu \rho}+R^\beta _{~\rho \mu \nu})S_{\beta \sigma}+{1\over 3}(R^\beta _{~\nu \mu \sigma}+R^\beta _{~\sigma\mu \nu})S_{\rho \beta}}_{Naive~covariantization} + (curvature \times stuckelberg~fields)
\ee

The Stuckelberg fields $S_{11\mu} ,S_{11}$ are required for gauge invariance. The gauge transformations are given in \eqref{GT2}.  They can be set to zero by a gauge transformation.

These expressions can now be substituted in \eqref{EOMspin2} and we obtain the free gauge invariant spin 2 equations in an arbitrary curved space-time background. Since it is tedious and the final answer is not especially illuminating we do not do this here. The final result is given in  Appendix B.

\subsection{Interacting Open String: Level 2}

The interaction terms ("gauge invariant field strengths"\cite{BSERGopen1,BSERGopen2}) expressed in terms of loop variables can be mapped to space-time fields using the same three step procedure. A new feature that enters is that the interaction involves a product of two such field strengths at {\em different points} on the world sheet. They are of the form
\be  \label{Int}
\int dz_1 dz_2 \dot G(z_1,z_2,a)K_{1\mu \nu \rho} [k_n]e^{i\ko.\bar Y(z_1)} \bar Y_1^\mu \bar Y_1^\nu \bar Y_1^\rho(z_1)K_{2\al \beta \gamma} [k'_n]e^{ik_0'.\bar Y(z_2)} \bar Y_1^\al \bar Y_1^\beta \bar Y_1^\gamma(z_2)
\ee
The exponentials have to be understood as a power series that stands for the Taylor expansion described in \eqref{Taylor}. Since one is not likely to have an exact expression for $G(z_1,z_2,a)$ in general backgrounds,  except as a power series in $z_1-z_2$, one also has to perform an operator product expansion (OPE) of the product of vertex operators in powers of $z_1-z_2$. 

In mapping these expressions to space-time fields the same three step procedure can be followed.  The point of departure being that each interaction product will involve an infinite series of terms involving higher derivatives
from the expansion of the exponentials, as well as the subsequent expansion in powers of $z_1-z_2$ involving  higher level operators.

Thus for the OPE of normal ordered exponentials in flat space we have:
\br  \label{OPE}
: e^{i\ko.Y(z_1)}::e^{ip_0.Y(z_2)}: &=& e^{-\kom p_{0\nu} \lan Y^\mu(z_1) Y^\nu(z_2)\ran}:e^{i\ko.Y(z_1)+ip_0.Y(z_2)}:\nonumber \\
&=&e^{-\kom p_{0\nu} \lan Y^\mu(z_1) (Y^\nu(z_1)+ (z_2-z_1)\p_z Y^\nu(z_1) + {(z_1-z_2)^2\over 2!} \pp _zY^\nu(z_1)+...\ran}\nonumber \\
& &:e^{i\ko.Y(z_1)+ip_0.(Y^\nu(z_1)+ (z_2-z_1)\p_z Y^\nu(z_1) + {(z_1-z_2)^2\over 2!} \pp _zY^\nu(z_1)+...)}:
\er

In curved space we use RNC $\bar Y^\mu$. We can further choose the origin of the coordinate system so that $\bar Y^\mu(z_1)=0$. We then expand the exponentials in powers of $p_0^\mu$ and derivatives of $\bar Y^\mu(z_1)$. Then the powers of $p_0^\mu$ will represent the Taylor series expansion \eqref{Taylor}. 
We give a few examples below: (We let $Y_o$ stand for the coordinates, in a general coordinate system $Y$, of the point that is the origin, i.e. $\bar Y=0$, of the RNC $\bar Y$, and bars denote the RNC.)
\br  \label{example}
\lan p_{0\mu} K_{\alpha \beta \gamma} [p_0,p_n]\ran &= & \bar \nabla _\mu F_{\alpha \beta \gamma} (\bar Y=0)\to \nabla _\mu F_{\alpha \beta \gamma} (Y_o)\\
\lan p_{0\mu}p_{0\nu} K_{\alpha \beta \gamma} [p_0,p_n]\ran &= & \bar \nabla _\mu \bar \nabla _\nu F_{\alpha \beta \gamma} (\bar Y=0)+{1\over 3}\Big((\bar R^\lambda_{~\alpha \mu \nu}+\bar R^\lambda_{~\nu \mu \alpha})F_{\lambda \beta \gamma}(0) +\nonumber \\ & &(\bar R^\lambda_{~\beta \mu \nu}+\bar R^\lambda_{~\nu \mu \beta})F_{\alpha \lambda \gamma}(0)+(\bar R^\lambda_{~\gamma \mu \nu}+\bar R^\lambda_{~\nu \mu \gamma})F_{\alpha \beta \lambda}(0)\Big)\nonumber \\&\to& \nabla _\mu \nabla _\nu F_{\alpha \beta \gamma} (Y_o)+{1\over 3}\Big((R^\lambda_{~\alpha \mu \nu}+R^\lambda_{~\nu \mu \alpha})F_{\lambda \beta \gamma}(Y_0) +\nonumber \\ & &(R^\lambda_{~\beta \mu \nu}+R^\lambda_{~\nu \mu \beta})F_{\alpha \lambda \gamma}(Y_0)+(R^\lambda_{~\gamma \mu \nu}+R^\lambda_{~\nu \mu \gamma})F_{\alpha \beta \lambda}(Y_0)\Big)
\er
All these expressions are tensors at the origin $Y_0$ of the RNC.
Also for the contractions one needs Taylor expansions of the Green function, for example
\be \label{contr}
\lan \bar Y^\mu (z_1)\bar Y^\nu(z_1)\ran = \eta^{\mu\nu} G(z_1,z_1;a) ~~~;~~~\lan \bar Y^\mu (z_1)\p_z \bar Y^\nu(z_1)\ran = \eta^{\mu\nu} \p_{z_2}G(z_1,z_2;a)|_{z_2=z_1}
\ee
In a general coordinate system we simply replace $\eta ^{\mu \nu}$ by $g^{\mu \nu}(Y_0)$ in the above equation.

Putting all this together one obtains on expanding the exponentials in \eqref{OPE}  
\[
\lan -\int \int~dz_1 dz_2~[\dot G(z_1,z_1;a)+ (z_1-z_2)\dot G'(z_1,z_1;a)+...][ 1-\kom p_{0\nu} g^{\mu\nu}(Y_0)G(z_1,z_1;a)+...]
\]
\[
K_{\lambda\sigma\rho}[\ko,\kn]\p_{z_1}Y^\lambda (z_1)\p_{z_1}Y^{\sigma}(z_1)\p_{z_1}Y^\rho(z_1)
K_{\alpha \beta \gamma}[p_0,p_n]\p_{z_1}Y^\alpha(z_1)\p_{z_1}Y^\beta(z_1)\p_{z_1}Y^\gamma(z_1)
\]
\[
e^{i(\ko+p_0).Y_0} (1+ (z_2-z_1)p_{0}.\p_{z_1}Y(z_1)+...)\ran
\]
\[
=-\int \int~dz_1 dz_2~[\dot G(z_1,z_1;a)+ (z_2-z_1)\dot G'(z_1,z_1;a)+...]\]\[[F_{\lambda \sigma\rho}(Y_0) F_{\alpha \beta \gamma}(Y_0)- g^{\mu\nu}(Y_0) \nabla_\mu F_{\lambda \sigma\rho}(Y_0)\nabla_\nu F_{\alpha \beta \gamma}(Y_0)G(z_1,z_1;a)+...]
\]\be
\p_{z_1}Y^\lambda (z_1)\p_{z_1}Y^{\sigma}(z_1)\p_{z_1}Y^\rho(z_1)
\p_{z_1}Y^\alpha(z_1)\p_{z_1}Y^\beta(z_1)\p_{z_1}Y^\gamma(z_1)+...
\ee

where we have kept a sample term at level 6. There are also terms at lower levels involving further contractions, and also terms at higher levels coming from expanding the exponential.

\subsection{Closed String: Level $(2,\bar 2)$}

As we have seen, the method works in almost exactly the same way for free or interacting case. For the interacting case we need to write each term as a Taylor expansion and each term in the expansion is mapped using the same methods. Let us look at the free equation here.
The free equation of motion (EOM) for the four index tensor field in terms of loop variables can be written as \cite{BSGIC,BSERGclosed}:
\[
   -{1\over 4} \ko^2 (\ki .Y_1)^2 (k_{\bar 1} .Y_{\bar 1})^2 +\hf \ko.\ki (\ko .Y_1)(\ki .Y_1) (\kib .\yib)^2 + \hf \ko.\kib (\ko.\yib)(\kib .\yib)(\ki.\yi)^2 +
\]
\be
   -{1\over 4} \ki.\ki (\ko.\yi)^2(\kib.\yib)^2 -{1\over 4} \kib.\kib (\ko.\yib)^2(\ki.\yi)^2 - \ki.\kib (\ko.\yi)(\ko.\yib)(\ki.\yi)(\kib.\yib)
\ee
  It is gauge invariant under 
\[
  \kim \to \kim + \li \kom;~~~~\kimb \to \kimb + \la_{\bar1} \kom
\] 
  if we use the tracelessness condition on the gauge parameters: 
\be   \label{const}
   \li \ki.\kib \kimb= \li \kib.\kib \kim =0= \lib \ki.\kib \kim = \lib \ki.\ki \kimb
\ee

The fields are also defined in \cite{BSGIC}.
\br
\lan \kim \kin \kirb \kisb \ran &=&S_{1\mu 1\nu \bar 1\rho \bar 1\sigma }\nonumber \\
\lan q_1  \kin \kirb \kisb \ran &=& S_{11\nu\bar 1\rho \bar 1\sigma }\qo \nonumber \\
\lan q_{\bar 1} \kim \kin \kirb \ran&=& S_{1\mu 1\nu \bar 1 \rho 1}\qo \nonumber \\
...& & \\
\er
and similarly for the remaining fields.  We hope the notation is clear to the reader.
Consider the first term (the bar on $Y$ indicates RNC): 
\be
 \ko^2 (\ki .\bar Y_1)^2 (k_{\bar 1} .\bar Y_{\bar 1})^2 = \ko^2 \kim \kin k_{\bar 1 \rho} k_{\bar 1 \sigma} \bar Y_1^\mu \bar Y_1^\nu \bar Y_{\bar 1}^\rho \bar Y_{\bar 1}^\sigma
\ee
Thus we need to map $ \ko^2 \kim \kin \kirb \kisb$ to a space-time field using our modified prescription. 

The four index tensor equation map is quite tedious to work out. There is no new complication that arises except that we need the Taylor expansion in RNC \eqref{Taylor} to higher orders. So we will only give outlines. 

The constraints \eqref{const}  can be mapped directly to space-time field constraints. If it is zero in flat space, it continues to be zero even in curved space since the extra curvature couplings in curved space are also linear in the constraint.

{\bf Step 1}

Let $\kim = \tkim + y_1 \kom$ and $\kimb = \tilde k_{\bar 1\mu} + y_{\bar 1} \kom$. Then we obtain:
\be
\ko^2 \kim \kin \kirb \kisb= \ko^2(\tkim +y_1\kom)(\tkin y_1\kon)(\tilde k_{\bar 1\rho} + y_{\bar 1} \kor)(\tilde k_{\bar 1\sigma} + y_{\bar 1} \kos)
\ee

We define some tilde fields at the intermediate stage as:
\br
\lan \tkim \tkin \tilde k_{\bar 1\rho} \tilde k_{\bar 1\sigma} \ran &=&\tilde S_{1\mu 1\nu \bar 1\rho \bar 1\sigma}\nonumber \\
\lan y_1  \tkin   \tilde k_{\bar 1\rho} \tilde k_{\bar 1\sigma}\ran &=& \tilde S_{11\nu \bar 1\rho\bar 1 \sigma} \nonumber \\
\lan y_{\bar 1}\tkim  \tkin   \tilde k_{\bar 1\rho} \ran &=& \tilde S_{1\mu 1\nu \bar 1\bar 1\rho } \nonumber \\
\lan y_1 y_{\bar 1} \tkim \tilde k_{\bar 1\sigma}\ran &=& \tilde S_{11\mu \bar 1\bar 1 \sigma}\nonumber \\
\lan y_{\bar 1}^2 \tkim \tkin\ran &=& \tilde S_{1\mu 1\nu \bar 1 \bar 1} \nonumber \\
...& &\nonumber\\
\er
etc.

We need to work out the map between these sets of fields.
\br   \label{2tensor}
\tilde S_{11\bar 1 \bar 1}&=&{1\over \qo^4}S_{11\bar 1 \bar 1}\nonumber \\
\tilde S_{11\mu \bar 1 \bar 1} &=&  {S_{11\mu \bar 1 \bar 1}\over \qo^3} - {\nabla_\mu S_{11\bar 1 \bar 1}\over \qo^4}\nonumber \\
\tilde S_{11 \bar 1\rho \bar 1\sigma}&=& {S_{11 \bar 1\rho \bar 1\sigma}\over \qo^2} - { \nabla_{(\rho}S_{11 \bar 1 \bar 1\sigma)}\over \qo^3}
+{ \nabla_\rho\nabla_\sigma S_{11\bar 1 \bar 1}\over \qo^4}
\er
The above equations are essentially the same as was given in the last section for open strings. We further need expressions for the three and four index tensors.

After some straightforward algebra one finds the following relation for the three index tensor:
\br   \label{3tensor}
\tilde S_{11\nu \bar 1 \rho \bar 1 \sigma}&= & {S_{11\nu \bar 1 \rho \bar 1 \sigma}\over \qo} -{1\over \qo^2}[ \nabla_\nu S_{11\bar 1 \rho \bar 1 \sigma}+\nabla_\rho S_{11\nu \bar 1  \bar 1 \sigma}+\nabla_\sigma S_{11\nu \bar 1 \rho \bar 1 }]\nonumber \\ 
& &+ {1\over \qo^3}[\nabla_{\rho} \nabla_\nu S_{11 \bar 1 \bar 1\sigma} +\nabla_{\sigma} \nabla_\nu S_{11 \bar 1 \bar \rho1} +\nabla_{\sigma} \nabla_\rho S_{11\nu \bar 1 \bar 1}] + {1\over \qo^4}\nabla_\sigma \nabla _\nu \nabla_\rho  S_{11\bar 1 \bar 1}
\nonumber \\
& &+{2\over 3}(R^\lambda_{~\rho \nu \sigma}+R^\lambda_{~\sigma \nu \rho})[{S_{11 \bar 1 \bar 1\lambda}\over \qo^3}
-{\nabla_\lambda S_{11\bar 1 \bar 1}\over \qo^4}] +{1\over 3}(R^\lambda_{~\sigma\rho \nu }+R^\lambda_{~ \nu \rho \sigma})[{S_{11\lambda \bar 1 \bar 1}\over \qo^3}] 
\er

Finally the four index tensor satisfies a relation of the form
\[
S_{1\mu 1\nu \bar 1 \rho \bar 1 \sigma} = \tilde S_{1\mu 1\nu \bar 1 \rho \bar 1 \sigma} + (lower~index ~tensors)
\]

Using \eqref{2tensor} and \eqref{3tensor}, one can solve for $ \tilde S_{1\mu 1\nu \bar 1 \rho \bar 1 \sigma}$ in terms of the ordinary fields.
We do not work it out here.

{\bf Step 2}

Using the results of \eqref{Taylor} we obtain for instance:
\br    \label{cl-step2}
\lan \ko^2 \tkim \tkin \tilde k_{\bar 1\rho}\tilde k_{\bar 1\sigma}\ran &=&\nabla^2 \tilde S_{11\bar 1\bar 1 \mu \nu \rho \sigma}-{1\over 3}(R^\lambda _{~\mu} \tilde S_{11\bar 1\bar 1 \lambda \nu \rho \sigma}+R^\lambda _{~\nu} \tilde S_{11\bar 1\bar 1 \mu \lambda  \rho \sigma}+R^\lambda _{~\rho} \tilde S_{11\bar 1\bar 1 \mu \nu \lambda \sigma}+R^\lambda _{~\sigma} \tilde S_{11\bar 1\bar 1 \mu \nu \rho \lambda})\nonumber \\
\lan\ko^2 y_1 \kom \tkin \tilde k_{\bar 1\rho}\tilde k_{\bar 1\sigma}\ran &=&\nabla_\mu \nabla^2 \tilde S_{11\bar 1\bar 1 \nu \rho \sigma}-(R^\lambda _{~\nu}\nabla_\mu \tilde S_{11\bar 1\bar 1 \lambda \rho \sigma}+R^\lambda _{~\rho}\nabla _\mu \tilde S_{11\bar 1\bar 1\nu  \lambda  \sigma}+R^\lambda _{~\sigma}\nabla _\mu \tilde S_{11\bar 1\bar 1 \nu \rho \lambda}) \nonumber \\ & &- \hf (\nabla _\mu R^\lambda _{~\nu} \tilde S_{11\bar 1\bar 1\lambda  \rho  \sigma}+\nabla _\mu R^\lambda _{~\rho} \tilde S_{11\bar 1\bar 1\nu  \lambda  \sigma}+\nabla _\mu R^\lambda _{~\sigma} \tilde S_{11\bar 1\bar 1 \nu \rho \lambda})
\er
We do not bother to write down the rest of the terms. As the number of derivatives increase the expressions become more complicated. Hopefully it is clear to the reader that given the taylor expansion \eqref{Taylor} the terms can easily be written down.
 
{\bf Step 3}

The last step is to plug in the results of step 1 into \eqref{cl-step2}. It should be clear that all higher derivative terms not involving the curvature tensor cancel and reproduce the flat space result. Then the curvature couplings to the four index tensor field give the naive covariantization just as in \eqref{op-step3}.  The remaining terms are curvature coupling to Stuckelberg fields. These are required for gauge invariance and can be set to zero by a choice of gauge. 

This concludes our outline of the description of how space-time field equations are obtained from loop variable expressions. It works for closed and open strings in exactly the same way. It is a well defined construction (albeit tedious). The final result is an expression that is 
generally covariant as well as gauge invariant.

\section{Covariance of World Sheet Action}

In the last two sections we showed how one obtains covariant {\em and} gauge invariant equations of motion. In this section for logical completeness we show that the world sheet action that one starts out with can easily be written in a  covariant form with vertex operators  written in terms of covariant derivatives such that in  RNC they reduce to the ones we have been working with. The main observation is that vertex operators of the form $\p^3 \bar Y^\mu\over \p x_p \p x_n \p x_m$ although written in terms of ordinary derivatives can be understood as covariant tensors at the origin of the RNC.

\subsection{Some properties of RNC}

We recollect some basic results about  RNC \cite{AGFM,Pet,Eis}.
\footnote{Some properties are also given in the Appendix A}
The geodesic equation is
\be   \label{geodesic}
{d^2 X^a\over d \tau^2} + \Gamma ^a_{bc} \dot X^b \dot X^c =0
\ee

In RNC ${d^2 \bar Y^a\over d \tau^2}=0$. Therefore 
\be
\bar \Gamma ^a_{bc} \dot {\bar Y}^b \dot {\bar Y}^c=0
\ee

At the origin $\dot {\bar Y}^\mu$ can point in any direction. So we conclude that 
\be
\bar \Gamma ^a_{bc}(0)=0
\ee
Differentiating \eqref{geodesic} we get
\br
{d^3 X^a \over d \tau^3} &=& - {d\over d\tau}[\Gamma ^a_{bc} \dot X^b \dot X^c] \nonumber \\
&=& - \p_d \Gamma ^a_{bc} \dot X^d \dot X^b \dot X^c - \Gamma ^a_{bc} {d\over d\tau}[  \dot X^b \dot X^c]\nonumber \\
&=&  - \p_d \Gamma ^a_{bc} \dot X^d \dot X^b \dot X^c-\Gamma ^a_{bc}[ -\Gamma ^b_{de}\dot X^d \dot X^e] \dot X^c - \Gamma ^a_{bc}\dot X^b [ - \Gamma^c_{de} \dot X^d \dot X^e] \nonumber \\
&=& -\underbrace{[ \p_d \Gamma ^a_{bc}- \Gamma ^a_{bi}\Gamma^i_{dc} - \Gamma^a_{ic}\Gamma^i_{db}]}_{\Gamma^a_{bcd}}\dot X^d \dot X^b \dot X^c
\er
Thus 
\be
{d^3 X^a \over d \tau^3}=-\Gamma^a_{bcd}\dot X^d \dot X^b \dot X^c= -{1\over 3!}\Gamma^a_{(bcd)}\dot X^d \dot X^b \dot X^c\equiv \tilde \Gamma^a_{bcd}\dot X^d \dot X^b \dot X^c
\ee
Using the manifest symmetry in $b,c,d$ we have defined in the above equation, a symmetric tensor $\tilde \Gamma^a_{bcd}$.
Once again using ${d^3 \bar Y^a \over d \tau^3}=0$ in the RNC one obtains that ${\tilde {\bar \Gamma}^a}_{bcd}(0)=0$.

This pattern continues recursively and one obtains at the next level
\be
{d^4X^a\over d\tau^4}=-[\p_e \tilde \Gamma^a_{bcd} - \tilde \Gamma^a_{icd}\Gamma^i_{eb}-\tilde \Gamma^a_{bid}\Gamma^i_{ec}-\tilde \Gamma^a_{bci}\Gamma^i_{ed}]\dot X^e\dot X^d \dot X^b \dot X^c
\ee
Thus
\be
{\tilde {\bar \Gamma}^a}_{bcde}(0)={1\over 4!} [\p_e \tilde{\bar \Gamma}^a_{bcd} - \tilde {\bar \Gamma}^a_{icd}{\bar \Gamma}^i_{eb}-\tilde {\bar \Gamma}^a_{bid}{\bar \Gamma}^i_{ec}-\tilde {\bar \Gamma}^a_{bci}\bar{\Gamma}^i_{ed}]_{symmetrized~on~bcde}(0)=0
\ee
and so on.

Now it can be shown that it is precisely this combination of $\Gamma$ matrices that occurs in vertex operators {\em provided we symmetrize}.

\subsection{Vertex Operators}

We start with the vector ${\p Y^\mu \over \p x^\alpha} \equiv \p_\alpha Y^\mu \equiv  Y^\mu_\alpha$ where $x^\alpha$ is some parameter on the world sheet. It could stand for the world sheet coordinate $z,\bar z$ or the loop variable coordinates $\xn$.
A covariant derivative was defined in \cite{BSERGclosed,BSGIC}:
\be
D_\beta Y^i_\alpha \equiv \p_\beta Y^i_\alpha+\Gamma ^i_{ba}Y^b_\beta Y^a_\alpha 
\ee
Then
\be
\p_\beta Y^i_\alpha = D_\beta Y^i_\alpha - \Gamma ^i_{ba}Y^b_\beta Y^a_\alpha
\ee
 Clearly in RNC 
\be
\p_\beta \bar Y^a = D_\beta \bar Y^a
\ee

Now consider the next derivative
\br
\p_\gamma \p_\beta Y_\alpha ^i &=&D_\gamma D_\beta X_\alpha ^i- \Gamma^i_{ca}Y^c_\gamma D_\beta Y^a_\alpha - \p_\gamma [\Gamma ^i_{ba} Y^b_\beta Y^a_\alpha] \nonumber \\
&=&D_\gamma D_\beta X_\alpha ^i- \Gamma^i_{ca}Y^c_\gamma D_\beta Y^a_\alpha -(\p_\gamma \Gamma ^i_{ba}) Y^b_\beta Y^a_\alpha
-\Gamma ^i_{ba} \p_\gamma[Y^b_\beta Y^a_\alpha]\nonumber \\
&=&D_\gamma D_\beta X_\alpha ^i- \Gamma^i_{ca}Y^c_\gamma D_\beta Y^a_\alpha -(\p_\gamma \Gamma ^i_{ba}) Y^b_\beta Y^a_\alpha
-\Gamma ^i_{ba}D_\gamma[Y^b_\beta Y^a_\alpha]-\Gamma ^i_{ba}[-\Gamma^b_{cd}Y_\gamma^c Y_\beta ^d Y_\alpha^a - \Gamma ^a_{cd}
Y^c_\gamma Y^b_\beta Y^d_\alpha ]\nonumber\\
&=&D_\gamma D_\beta Y_\alpha ^i -\Gamma^i_{ca}Y^c_\gamma D_\beta Y^a_\alpha-\Gamma ^i_{ba}D_\gamma[Y^b_\beta Y^a_\alpha]-(\p_c \Gamma ^i_{ba})Y^c_\gamma Y^b_\beta Y^a_\alpha-\Gamma ^i_{ba}[-\Gamma^b_{cd}Y_\gamma^c Y_\beta ^d Y_\alpha^a - \Gamma ^a_{cd}
Y^c_\gamma Y^b_\beta Y^d_\alpha ]\nonumber\\
&=&D_\gamma D_\beta Y_\alpha ^i -\Gamma^i_{ca}Y^c_\gamma D_\beta Y^a_\alpha-\Gamma ^i_{ba}D_\gamma[Y^b_\beta Y^a_\alpha]
-\Gamma^i_{bac}Y^c_\gamma Y^b_\beta Y^a_\alpha \nonumber
\er 
We can now symmetrize the RHS in $\alpha, \beta , \gamma$ because the LHS is symmetric, and write
\be
\p_\gamma \p_\beta Y_\alpha ^i={1\over 3!}\{D_{(\gamma} D_\beta Y_{\alpha )} ^i -\Gamma^i_{ca}Y^c_{(\gamma} D_\beta Y^a_{\alpha )}-\Gamma ^i_{ba}D_{(\gamma}[Y^b_\beta Y^a_{\alpha )}]
-\tilde \Gamma^i_{bac}Y^c_{( \gamma} Y^b_\beta Y^a_{\alpha )}\}
\ee 
We have used the fact that $Y^c_{( \gamma} Y^b_\beta Y^a_{\alpha )}$ is also symmetric now in $a,b,c$.  It is clear from the above pattern that symmetrized vertex operators will involve the $\tilde \Gamma^a_{bcde...}$ as defined in the last subsection.

If we now specialize to RNC we find that
\be
\p_\gamma \p_\beta \bar Y_\alpha ^i={1\over 3!}\{D_{(\gamma} D_\beta \bar Y_{\alpha )}^i
\ee
We will thus take the RHS as the definition of our vertex operator and use it in a general coordinate system. Thus for instance,
a term 
\be
K_{mnp\mu}\bar Y^\mu_{mnp}= K_{mnp\mu} {1\over 3!}D_{(m}D_nD_{p)} Y^\mu
\ee
 is manifestly invariant if we take $K_{mnp\mu}$ to be a vector. There is a subtlety here due to the presence of the factor $e^{i\ko.Y}$. This means that \be 
 \lan K_{mnp\mu}e^{i\ko.Y}\ran = S_{mnp\mu}(Y)
 \ee
 $S_{mnp\mu}(Y)$ is a vector at $Y$, not at the origin. But we can use \eqref{Taylor} to Taylor expand it in the RNC as a sum of vectors at the origin. Multiplying by ${1\over 3!}D_{(m}D_nD_{p)} Y^\mu$ gives a sum of scalars at the origin. Being a scalar it has the same value in any coordinate system. Thus we have an expression for a coordinate invariant action, that reduces to the required action in the RNC.
 We can take this as the action in a general coordinate system. We can then choose RNC without any loss of generality and work in that system. In this coordinate system we know (and checked by explicit construction) that the ERG equations are gauge invariant. Thus we are justified in working out the equations in this coordinate system and then covariantizing in the usual way - as was done in the last two sections.

\section{Conclusion}

In this paper we have obtained gauge invariant and generally covariant equations for massive higher spin fields of string theory using the ERG. The background space-time can have any metric and the method is not tied to any specific choice. In this sense the method is {\em manifestly} background independent. 
 
   The main ingredient in this paper is a new prescription for mapping from the loop variable equation  to space time field equations.
   
There are many open questions. A few are listed below:
\begin{enumerate}
\item It would be interesting to construct an action.

\item It would be interesting to obtain some non trivial solutions of the exact RG. 

\item In \cite{BSGIC} it was shown that the field strength for the graviton  can be made gauge invariant for non zero mass quite easily. However the constraint of zero mass forced us to modify the gauge transformation to include coordinate transformations. This made the kinetic term non invariant and gave extra contributions, and ultimately a gauge invariant field strength. If one attempts this for the massive modes, this should give some insight into some more symmetrical phase of string theory where all modes are massless.

\item It would be interesting to pursue the speculations in \cite{BSLV} regarding the connection between symmetries of string theory and the space-time renormalization group.
\end{enumerate}   
 
We hope to return to these questions.

\section{Appendix A}
We list a few useful formulae that can be obtained from \eqref{Taylor} and the basic commutation rule
\br
[\nabla _\mu, \nabla_\nu] S_{\al_1} &=& -R^\beta _{~\al_1 \mu \nu} S_\beta\\
{\pp S_{\al_1\al_2}\over \p y^\mu \p y^\nu}|_{y=0} &=&\hf[ \nabla _{(\mu }\nabla _{\nu)}S_{\al_1 \al_2} + {1\over 3} R^\beta_{~ (\mu \nu) \al_1}S_{\beta \al_2} +R^\beta_{~ (\mu \nu) \al_2}S_{ \al_1\beta}]\nonumber \\
&=& \nabla _\mu \nabla_\nu S_{\al_1 \al_2} + {1\over 3} (R^\beta_{~\nu \mu \al_1}+R^\beta_{~\al_1 \mu \nu}) S_{\beta \al_2}+{1\over 3} (R^\beta_{~\nu \mu \al_2}+R^\beta_{~\al_2 \mu \nu}) S_{ \al_1 \beta}\nonumber \\
{\ppp S \over \p y^\mu \p y^\nu \p y^\rho}|_{y=0} &=&{1\over 3!} \nabla _{(\rho} \nabla _\nu \nabla_{\mu)}S\nonumber \\& =&
\nabla _\rho \nabla _\nu \nabla_\mu S + {1\over 3} (R^\beta _{~\nu \rho \mu}+ R^\beta_{~\mu \rho \nu})\nabla_\beta S\nonumber
\\
{\ppp S_{\al_1}\over \p y^\mu \p y^\nu \p y^\rho}|_{y=0} &=& {1\over 3!} [\nabla_{(\rho}\nabla_\nu \nabla_{\mu)}S_{\al_1} - R^\beta_{~(\mu|\al_1|\nu}\nabla_{\rho)}S_\beta -\hf \nabla_{(\rho}R^\beta_{~\mu|\al_1|\nu)}S_\beta ]\nonumber
\\
{\p^4 S \over \p y^\mu \p y^\nu \p y^\rho \p y^\sigma}|_{y=0} &=&{1\over 4!} \nabla _{(\rho}\nabla_\sigma \nabla _\nu \nabla_{\mu)}S \nonumber \\&=& \nabla_\mu \nabla_\nu \nabla_\rho \nabla_\sigma S +[{1\over 4} (R^\beta _{~\nu \mu \rho} \nabla_\sigma \nabla_\beta + R^\beta _{~\sigma \mu \rho} \nabla_\nu \nabla_\beta + R^\beta _{~\sigma \nu \rho} \nabla_\mu \nabla_\beta)S + (\rho \leftrightarrow \sigma)] \nonumber
\\
& &+ {1\over 4}(R^\beta _{~\rho \mu \nu} \nabla_\sigma \nabla_\beta + R^\beta _{~\sigma \mu \nu} \nabla_\rho \nabla_\beta)S + [{1\over 12}(R^\beta _{~\sigma\mu \nu} +R^\beta _{~\nu\mu \sigma})\nabla_\rho \nabla_\beta S + (\rho \leftrightarrow \sigma)]\nonumber
\\
& &+[{1\over 12} (R^\beta _{~\rho \nu \sigma}+R^\beta _{~\sigma\nu \rho)}\nabla_\mu \nabla _\beta S +(\mu \leftrightarrow \nu )]\nonumber
\\
& &+ \hf (\nabla_\mu R^\beta_{~\sigma \nu \rho} + (\nabla_\mu R^\beta_{~\rho \nu \sigma})\nabla_\beta S +
[{1\over 12} \nabla_\mu ( R^\beta_{~\sigma \nu \rho}+R^\beta_{~\rho \nu \sigma})\nabla_\beta S + (\mu \leftrightarrow \nu )]\nonumber
\\
& &+[{1\over 12}\nabla_\rho (R^\beta_{~\sigma \mu \nu }+R^\beta_{~ \nu \mu \sigma})\nabla_\beta S + (\rho \leftrightarrow \sigma)]
\er
\section{Appendix B}

The basic commutation rule for covariant derivatives
\[
[\nabla_\mu ,\nabla_\nu] S_\alpha = - R^\beta_{~\alpha \mu \nu}S_\beta
\]
can be used to simplify the derivative terms. In this Appendix, for simplicity we let $S_{11\mu \nu}=S_{\mu\nu},S_{11\mu}=S_\mu,S_{11}=S$. We define the following tensor:
\br
\lan \kom \kon \kir \kis \ran &=& \hf [\nabla_{(\mu}\nabla_{\nu)} S_{\rho \sigma} -{1\over 3} R^\beta_{~(\mu |\rho|\nu)}S_{\beta \sigma}-{1\over 3}R^\beta_{~(\mu |\sigma|\nu)}S_{\rho \beta} ] + \nonumber \\ & &{1\over 6}[R^\beta_{~(\mu |\rho|\nu)}({\nabla_{(\beta}S_{\sigma)}\over \qo} -{\nabla_\beta \nabla_\sigma S\over \qo^2}) +R^\beta_{~(\mu |\sigma|\nu)}({\nabla_{(\beta}S_{\rho)}\over \qo} -{\nabla_\beta \nabla_\rho S\over \qo^2})]-\nonumber \\
& & {1\over 6}[R^\beta_{~(\mu |\sigma|\nu}\nabla_{\rho)}({S_\beta\over \qo} - {\nabla _\beta S\over \qo^2}) -\hf \nabla_{(\rho}R^\beta_{~\mu |\sigma|\nu)}({S_\beta\over \qo} - {\nabla _\beta S\over \qo^2})]\nonumber \\
& &-\hf[3 R^\beta_{~\sigma \rho \nu}\nabla_\mu S_\beta +3 R^\beta_{~\sigma \rho \mu}\nabla_\nu S_\beta + R^\beta_{~\nu \rho \mu}\nabla_\beta S_\sigma +R^\beta_{~\mu \rho \nu}\nabla_\beta S_\sigma + 3 R^\beta_{~ \rho \nu \mu}\nabla_\beta S_\sigma+3 R^\beta_{~ \sigma \nu \mu}\nabla_\rho S_\beta]\nonumber \\
& &+2(\nabla_\nu R^\beta_{~ \sigma \rho  \mu})S_\beta+2(\nabla_\mu R^\beta_{~ \sigma \rho  \nu})S_\beta]+\hf[R^\beta_{~ \sigma \nu  \mu}\nabla _\rho S_\beta +R^\beta_{~ \rho \nu  \mu}\nabla _\beta S_\sigma] \nonumber \\
& \equiv& F_{\mu\nu\rho\sigma}
\er

The EOM \eqref{EOMspin2mass} can now be written in terms of $F$ as 
\be
G^{\alpha \beta}[F_{\alpha \beta \mu \nu} - F_{\alpha(\mu|\beta|\nu)} + F_{\mu \nu \alpha \beta}] + \qo^2 S_{\mu\nu} -\qo \nabla_{(\mu}S_{\nu)}+\nabla_\mu \nabla_\nu S=0
\ee
where $G_{\mu\nu}$ is the space-time metric.
Note that the Stuckelberg fields $S_\mu,S$ can be set to zero by a gauge transformation \eqref{GT2}.

\end{document}